\documentclass[11pt]{article}
\begin{document}

\def\slash#1{{\rlap{$#1$} \thinspace/}}

\begin{flushright} 
January, 2003 \\
\end{flushright} 

\vspace{0.1cm}

\begin{Large}
       \vspace{1cm}
  \begin{center}
   {On Higher Dimensional Fuzzy Spherical Branes }      \\
  \end{center}
\end{Large}
\vspace{1cm}

\begin{center}
{\large  Yusuke Kimura }  \\ 

\vspace{0.5cm} 
{\it Theoretical Physics Laboratory, \\
 RIKEN (The Institute of Physical and Chemical Research), }\\
{\it Wako, Saitama 351-0198, Japan} \\

\vspace{0.5cm} 

{kimuray@postman.riken.go.jp}

\vspace{0.8cm} 


\end{center}

 \vspace{1cm}

\begin{abstract}
\noindent
\end{abstract}
\noindent 
\hspace{0.4cm}

Matrix descriptions of 
higher dimensional spherical branes are investigated. 
It is known that a fuzzy $2k$-sphere is described by 
the coset space $SO(2k+1)/U(k)$ 
and has 
some extra dimensions. 
It is shown that a fuzzy $2k$-sphere 
is comprised of $n^{\frac{k(k-1)}{2}}$ 
spherical $D(2k-1)$-branes
and has 
a fuzzy $2(k-1)$-sphere at each point. 
We can understand the relationship 
between these two viewpoints by the 
dielectric effect. 
Contraction of the algebra is also discussed.


\newpage 
\section{Introduction}
\hspace{0.4cm}

Some new features of string theory appeared 
since the discovery of the $D$-brane \cite{pol}. 
One of the recent interesting developments 
is the appreciation of noncommutative geometry. 
A low energy effective action of $N$ coincident $D$-branes 
is described by the $U(N)$ Yang Mills theory. 
In this theory, $U(N)$ adjoint scalars represent 
the transverse coordinates of this system \cite{wittenbrane}. 
Since they are given by $U(N)$ matrices, this fact suggests that 
the space-time probed by $D$-branes is 
related to noncommutative geometry. 
It is discussed that 
a world volume theory on $D$-branes 
in the presence of a constant NS-NS two-form  
background is described by a 
noncommutative gauge theory \cite{SW,douglashull,chuho}. 
Myers showed that 
a low energy effective action of $N$ $D$-branes 
in the presence of a constant 
Ramond-Ramond background
has an additional Chern-Simons term, 
and 
the minimum of the potential is given by 
the configuration 
where transverse coordinates 
form a fuzzy sphere \cite{myers}. 
The ideas of noncommutative geometry found a 
prominent role in string theory.


Higher dimensional fuzzy spheres have been investigated in some 
contexts \cite{GroLP,castelino,holi,
guralnikramgoo,CMT,ramgoo,horamgooram,ramgoolamoddsphere}.  
The authors in \cite{horamgooram} showed that 
the matrix description of a fuzzy $2k$-sphere 
is given by an $SO(2k+1)/U(k)$ coset space 
which is a $k(k+1)$ dimensional space. 
The stabilizer group of this noncommutative manifold 
is not $SO(2k)$ but $U(k)$. 
For example, a fuzzy four-sphere 
is described by $SO(5)/U(2)$, being a six dimensional space. 
In general, a higher dimensional 
fuzzy sphere has some extra dimensions. 
The existence of them is 
important for the quantization of 
a higher dimensional sphere. 
It is well-known that 
noncommutative geometry is realized by the 
guiding center coordinates of electrons in 
a constant magnetic field. 
The fuzzy four-sphere is naturally realized 
by the quantum Hall system 
which was constructed in \cite{zhanghu}. 
The realization was first investigated 
in \cite{fabinger}. 
The system in \cite{zhanghu} is composed of particles 
moving on a four dimensional 
sphere under an $SU(2)$ gauge field. 
It was shown that the configuration space of this system is 
locally $S^{4}\times S^{2}$ which is consistent 
with the result in \cite{horamgooram}. 
There were further analyses in 
\cite{chenhou,karanair,ykfoursphere,yxchen,masuda} 
following these papers. 

The purpose of this paper is to investigate 
higher dimensional spherical branes 
in matrix models 
and to elucidate the extra dimensions 
which appear from fuzzy spheres. 
The organization of this paper is as follows. 
A fuzzy four-sphere is considered as a example 
of higher dimensional fuzzy spheres 
in section \ref{sec:Fuzzyfour-sphere}. 
Some parts overlap with \cite{ykfoursphere}. 
We begin with the $SO(5,1)$ algebra. 
It is explained that 
the manifold which is described by this algebra 
becomes locally $S^{4}\times S^{2}$ in a large $N$ limit. 
The extra two dimensional sphere is added 
to make a four-sphere 
symplectic manifold. 
The procedure of the In\"{o}n\"{u}-Wigner contraction 
gives a six dimensional noncommutative plane 
whose algebra is given by the Heisenberg algebra. 
Two out of the six dimensions come from the extra 
two-sphere. 
In section \ref{sec:Fuzzy2k-sphere} 
we study higher dimensional 
fuzzy spheres. A fuzzy $2k$-sphere is defined by 
the $SO(2k+1,1)$ algebra. 
It is shown that 
a fuzzy $2k$-sphere with a radius $\alpha^{2}n(n+2k)$ 
has a fuzzy $2(k-1)$-sphere with 
a radius $\alpha^{2}n(n+2k-2)$ at each point. 
A fuzzy $2k$-sphere is a $k(k+1)$ dimensional 
space including a extra $k(k-1)$ dimensional space 
which originates from a fuzzy $2(k-1)$-sphere. 
Taking a large $N$ limit leads to 
a smooth manifold which is given by 
$S^{2k}\times S^{2(k-1)} \times \cdots \times S^{2}$. 
Contraction of the algebra gives a $k(k+1)$ 
dimensional noncommutative plane. 
In section \ref{sec:fuzzysphereandmatrixmodel} 
we give some comments about matrix models 
whose classical solutions are fuzzy spheres. 
If we expand matrices around a fuzzy sphere 
solution, a matrix model action gives 
a noncommutative Yang-Mills action on it. 
The advantage of such formulation is that 
we can define the derivative 
and formulate matrix models 
in terms of field theories. 
The size of matrices plays a role of 
cut-off, and noncommutative Yang-Mills theories 
have no divergences. 
In 
section \ref{sec:Dielectriceffectandfuzzysphere} 
we evaluate the value of the matrix model action for 
fuzzy spheres. 
The estimation leads to the fact that 
a fuzzy $2k$-sphere is comprised of $n^{\frac{k(k-1)}{2}}$ 
$D(2k-1)$-branes in the matrix model. 
The dielectric effect of $D$-branes 
manifests the structure of a higher dimensional 
fuzzy sphere in an interesting way. 
Section \ref{sec:summaryanddiscussions} is 
devoted to summary and discussions.


\section{Fuzzy four-sphere}
\label{sec:Fuzzyfour-sphere}
\subsection{Matrix description of fuzzy four-sphere}
\hspace{0.4cm}
In this section, we briefly review the algebra of 
fuzzy four-sphere \cite{castelino,ramgoo,horamgooram}.
As a beginning, we examine the 
$SO(5,1)$ algebra which is given by 
\begin{equation}
[\hat{J}_{MN},\hat{J}_{OP}]=2\left(
\eta_{NO}\hat{J}_{MP}
+\eta_{MP}\hat{J}_{NO}
-\eta_{MO}\hat{J}_{NP}
-\eta_{NP}\hat{J}_{MO}
\right), 
\label{so(5,1)}
\end{equation}
where $M,N,\ldots$ run over $0$ to $5$ and 
$\eta_{MN}=diag(-1,1,\ldots,1)$. 
These matrices are constructed as 
\begin{eqnarray}
&&\hat{J}_{0\mu}=\left(
\Gamma_{\mu}\otimes 1\otimes \cdots \otimes 1
+1\otimes \Gamma_{\mu} \otimes \cdots \otimes 1
+\cdots 
+1\otimes 1\otimes \cdots \otimes \Gamma_{\mu}
\right)_{Sym}, \cr
&&\hat{J}_{\mu\nu}=\left(
\Gamma_{\mu\nu}\otimes 1\otimes \cdots \otimes 1
+1\otimes \Gamma_{\mu\nu} \otimes \cdots \otimes 1
+\cdots 
+1\otimes 1\otimes \cdots \otimes \Gamma_{\mu\nu}
\right)_{Sym}, 
\end{eqnarray}
where $\mu$ and $\nu$ run over $1$ to $5$. 
These are 
constructed from the 
$n$-fold symmetric tensor product of 
the five dimensional 
Gamma matrices and 
$Sym$ means the completely symmetrized tensor product. 
The size of the matrices is 
given by 
\begin{equation}
N=\frac{1}{6}(n+1)(n+2)(n+3). 
\label{size}
\end{equation}

We now explain that 
this algebra describes noncommutative geometry 
which becomes locally 
$S^{4}\times S^{2}$ 
in a large $N$ limit 
if we regard 
$J_{0\mu}=G_{\mu}$
\footnote{ 
Note that $\Gamma_{0\mu}$ is hermitian.}
 as coordinates of the four-sphere 
and $J_{\mu\nu}=G_{\mu\nu}$ 
as coordinates of the two-sphere. 
If we rewrite (\ref{so(5,1)}) using $G_{\mu}$ 
and $G_{\mu\nu}$, we have 
\begin{eqnarray}
&&
[\hat{G}_{\mu},\hat{G}_{\nu}] =
2\hat{G}_{\mu\nu}, \cr
&&[\hat{G}_{\mu},\hat{G}_{\nu\lambda}]=2\left(
\delta_{\mu\nu}\hat{G}_{\lambda}-\delta_{\mu\lambda}
\hat{G}_{\nu}
\right), \cr
&&[\hat{G}_{\mu\nu},\hat{G}_{\lambda\rho}]=2\left(
\delta_{\nu\lambda}\hat{G}_{\mu\rho}
+\delta_{\mu\rho}\hat{G}_{\nu\lambda}
-\delta_{\mu\lambda}\hat{G}_{\nu\rho}
-\delta_{\nu\rho}\hat{G}_{\mu\lambda}
\right), 
\label{fourspherealgebras}
\end{eqnarray}
where 
$\hat{G}_{\mu\nu}$ 
form the $SO(5)$ algebra. 
The fuzzy four-sphere is constructed to satisfy the 
following two conditions, 
\begin{equation}
\epsilon^{\mu\nu\lambda\rho\sigma}
\hat{x}_{\mu}\hat{x}_{\nu}\hat{x}_{\lambda}\hat{x}_{\rho}
=C\hat{x}_{\sigma} 
\label{foursphererelation}
\end{equation}
and
\begin{equation}
\hat{x}_{\mu}\hat{x}_{\mu}=\rho^{2}, 
\label{spherecondition}
\end{equation}
where $\hat{x}_{\mu}$ are coordinates 
of the fuzzy four-sphere and 
$\rho$ is a radius of the sphere. 
This sphere respects the $SO(5)$ invariance. 
Such $\hat{x}_{\mu}$ are constructed as 
\begin{equation}
\hat{x}_{\mu}=\alpha\hat{G}_{\mu},   
\end{equation}
where $\alpha$ is a dimensionful constant. 
If we use some formulae which are summarized 
in appendix \ref{sec:someformulae}, 
$C$ and $\rho$ is given by 
\begin{equation}
C=(8n+16)\alpha^{3} 
\end{equation}
and 
\begin{equation}
\rho^{2}=\alpha^{2}n(n+4). 
\end{equation}

Let us next observe that a fuzzy two-sphere exists at each point 
on the fuzzy four-sphere. 
We can always diagonalize a matrix $\hat{G}_{\mu}$ 
out of the five matrices. 
We diagonalize $\hat{x}_{5}=\alpha\hat{G}_{5}$ 
as in appendix \ref{sec:gammmamatrices}.  
The $SU(2)\times SU(2)$ algebra is constructed from 
the $SO(4)$ algebra, 
which is the stabilizer group of a usual four-sphere:  
\begin{equation}
[\hat{N}_{i},\hat{N}_{j}]=2i\epsilon_{ijk}\hat{N}_{k}, 
\hspace{0.3cm}
[\hat{M}_{i},\hat{M}_{j}]=2i\epsilon_{ijk}\hat{M}_{k},
\hspace{0.3cm} 
[\hat{M}_{i},\hat{N}_{j}]=0,  
\label{twosphereonfoursphere}
\end{equation}
where 
\begin{eqnarray}
\hat{N}_{1}&=&-\frac{i}{2}\left(\hat{G}_{23}-\hat{G}_{14}\right), 
\hspace{1.8cm}
\hat{M}_{1}=-\frac{i}{2}\left(\hat{G}_{23}+\hat{G}_{14}\right), \cr 
\hat{N}_{2}&=&\frac{i}{2}\left(\hat{G}_{13}+\hat{G}_{24}\right),  
\hspace{2.1cm}
\hat{M}_{2}=\frac{i}{2}\left(\hat{G}_{13}-\hat{G}_{24}\right), \cr 
\hat{N}_{3}&=&-\frac{i}{2}\left(\hat{G}_{12}-\hat{G}_{34}\right), 
\hspace{1.8cm}
\hat{M}_{3}=-\frac{i}{2}\left(\hat{G}_{12}+\hat{G}_{34}\right) .
\label{defofNandM}
\end{eqnarray}
These are compactly summarized as 
\begin{equation}
\hat{N}_{i}=-\frac{i}{4}\eta_{ab}^{i}
\hat{G}_{ab}, \hspace{0.4cm}
\hat{M}_{i}=-\frac{i}{4}\bar{\eta}_{ab}^{i}
\hat{G}_{ab}, 
\label{thooftsymbol}
\end{equation}
where $\eta_{ab}^{i}=\epsilon_{iab4}-\delta_{ia}\delta_{4b}
+\delta_{ib}\delta_{4a}$, and 
$\bar{\eta}_{ab}^{i}$ is obtained by changing the signs 
in front of the second and third terms of $\eta_{ab}^{i}$. 
The Casimir of each $SU(2)$ algebra 
at the north pole is evaluated as 
\begin{eqnarray}
\hat{N}_{i}\hat{N}_{i} 
&=&\frac{1}{4}\left(
n+G_{5}\right)\left( n+4+G_{5}\right)
=n(n+2),  \cr
\hat{M}_{i}\hat{M}_{i}
&=&\frac{1}{4}\left(
n-G_{5}\right)\left( n+4-G_{5}\right)
=0, 
\end{eqnarray}
where we have used $G_{5}=n$. 
Then we have a fuzzy two-sphere, 
which is given by the ($n+1$) dimensional representation 
of $SU(2)$, at the north pole. 
The radius of the fuzzy two-sphere is given by 
$\alpha^{2}n(n+2)$, being comparable with that 
of the fuzzy four-sphere. 
Since the fuzzy four-sphere has the $SO(5)$ symmetry, 
we can state that the fuzzy two-sphere, which is given by 
the ($n+1$) dimensional representation of $SU(2)$, 
exists at each {\it point} 
on the fuzzy four-sphere. 

We later consider this noncommutative manifold 
as a classical solution of a matrix model. 
Since we expand the matrices $A_{\mu}$ around 
the $SO(5)$ vector $\hat{G}_{\mu}$, 
it may be natural to regard the two-sphere 
as a internal space. 
In this sense, this noncommutative space 
is called 
fuzzy four-sphere. 
Let us comment on the stabilizer group of this manifold. 
It is known that a usual four-sphere is described as 
$SO(5)/SO(4)$ 
and the stabilizer group is $SO(4)$. 
The paper \cite{horamgooram} reported that 
the stabilizer group of the fuzzy four-sphere is not $SO(4)$ 
but $U(2)$ whose generators are $\hat{M}_{i}$ ($i=1,2,3$) 
and $\hat{N}_{3}$. 
This reason can be understood from the 
viewpoint of the deformation quantization. 
Since a usual four-sphere 
does not have the 
Poisson structure, 
it is difficult to quantize such a space. 
On the other hand a manifold 
which is described by 
$SO(5)/U(2)$ has 
the Poisson structure. 
Therefore we quantize 
$SO(5)/U(2) \simeq SO(5)/SO(4) \times SU(2)/U(1) 
\simeq S^{4}\times S^{2}$ instead of 
$SO(5)/SO(4) \simeq S^{4}$. 
It is not difficult to understand 
that the fuzzy two-sphere plays the role 
of the symplectic form. 
This picture becomes more manifest 
when we consider a contraction 
of the algebra, 
which is considered in 
section \ref{sec:contractionfoursphere}. 

The authors of \cite{castelino} 
constructed this fuzzy four-sphere as 
a longitudinal five-brane in the context of 
BFSS matrix model. 
They showed that 
$N$ and $n$ represents the number of $D$-particles and 
that of longitudinal five-branes respectively. 
(In section \ref{sec:fuzzysphereandmatrixmodel}, 
we regard this system as $D$-branes in IIB matrix model. )
Therefore the fuzzy four-sphere is composed of 
$n$ overlapping longitudinal five-branes.  
The existence of the fuzzy two-sphere is 
explained via the dielectric effect found 
by Myers \cite{myers}. 
If we notice a point on the overlapping five-branes, 
we have $n$ $D$-particles. 
A collection of $n$ $D$-particles 
expands into the fuzzy two-sphere 
as in (\ref{twosphereonfoursphere}). 
The expansion is along the extra dimensions. 

Since $SO(5)/U(2)$ is isomorphic to $SO(6)/U(3)$, 
we may use either algebra as the starting point 
of the algebra which defines 
the fuzzy four-sphere. 
We have used the latter algebra for later convenience. 
(The Wick rotation is needed.)
The stabilizer 
group of the $SO(5,1)$ algebra 
is $U(2,1)$ whose generators are 
$\hat{M}_{i}$, $\hat{N}_{3}$ and $\hat{G}_{\mu}$.

Let us investigate the lower dimensional brane 
charge in the four-branes
\footnote{
In this paper, 
we use the term ``four-brane" to refer to 
a four dimensional object. 
If we treat it in IIB matrix model, it 
means $D3$-brane.
}. 
A two-brane charge is given by 
$tr[\hat{x}_{\mu},\hat{x}_{\nu}]
\propto tr(\hat{N}+\hat{M})$ and it is vanishing 
at each point on the four-sphere. 
It appears after the contraction of the algebra 
(see (\ref{6dimplane1}) or (\ref{6dimplane2})).

Since we are now considering the $N \times N$ matrix, 
we have $N$ quanta in this system. 
The area occupied by the unit quantum 
on the four-sphere is 
\begin{equation}
\frac{\frac{8}{3}\pi^{2}\rho^{4} n}{N}
=\frac{16\pi^{2}\rho^{4}n}{(n+1)(n+2)(n+3)}, 
\end{equation}
where $8\pi^{2}\rho^{4}/3$ is the area of the four-sphere. 
This is a noncommutative scale on the four-sphere. 
Since there are $n$ quanta on the fuzzy two-sphere, 
the number of the quanta on the fuzzy four-sphere 
is $N/n$. 
There are $n \sim N^{\frac{1}{3}}$  
quanta at each point on the four-sphere 
which has 
$n^{2} \sim N^{\frac{2}{3}}$ points. 

A classical sphere is expected to be recovered 
in a large $N$ limit. 
We consider a large $N$ limit with 
the radius $\rho$ of the sphere fixed. 
In other words, it is 
an $\alpha \rightarrow 0$ limit 
with $\rho$ fixed .
We define coordinates $\hat{w}_{\mu\nu}$ as 
\begin{equation}
\hat{w}_{\mu\nu}\equiv i\alpha G_{\mu\nu}. 
\end{equation}
The coordinates of the four-sphere commute each other 
in this limit, 
\begin{equation}
[\hat{x}_{\mu},\hat{x}_{\nu}]=-2i\alpha\hat{w}_{\mu\nu}
 \simeq O (\alpha \rho) \rightarrow 0.
\end{equation}
The coordinates $\hat{x}_{\mu}$ and $\hat{w}_{\mu\nu}$ 
also become commuting matrices, 
\begin{equation}
[\hat{x}_{\mu},\hat{w}_{\nu\lambda}]=0, 
\hspace{0.4cm}
[\hat{w}_{\mu\nu},\hat{w}_{\lambda\rho}]=0 .
\end{equation}
Thus we have obtained the classical manifold of 
$S^{4} \times S^{2}$.


\subsection{Noncommutative field theory on fuzzy four-sphere}
\label{section:NCGT}
\hspace{0.4cm}
In this subsection, a noncommutative gauge field 
theory on the fuzzy four-sphere is investigated 
using a matrix model \cite{ykfoursphere}. 
It is known that 
matrix models provide the definition of noncommutative 
Yang-Mills, and 
the idea is to expand matrices around 
a classical solution \cite{Li,CDS,AIIKKT}. 
We expand the matrices as follows, 
\begin{equation}
A_{\mu}=\hat{x}_{\mu}+ \alpha \rho \hat{a}_{\mu} 
=\alpha\rho \left( \frac{1}{\rho}\hat{G}_{\mu}+ \hat{a}_{\mu} \right). 
\label{expansionmatrices}
\end{equation}
The details of the model are explained in 
section \ref{sec:fuzzysphereandmatrixmodel}. 
Note that we expanded only around $\hat{x}_{\mu}$. 
We do not consider fluctuations 
around $\hat{w}_{\mu\nu}$. 
Fields on a noncommutative sphere 
are expanded by 
noncommutative analogue of 
the spherical harmonics. 
In this case, such a noncommutative 
spherical harmonics is given in 
\cite{ramgoo}. 
The bases are classified by the $SO(5)$ representations 
and the matrices are expanded by the irreducible 
representations of $SO(5)$. 
The irreducible representation is characterized by the Young diagram. 
It is labeled by the row length $(r_{1},r_{2})$ 
in this case. 
Note that the representations with $r_{2}=0$ correspond to 
a classical four-sphere. 
Summing up the dimensions of all irreducible representations 
with the condition $n\geq r_{1} \geq r_{2}$ leads 
to the square of $N$ \cite{ramgoo}. 
We write the noncommutative 
spherical harmonics as 
$\hat{Y}_{r_{1}r_{2}}(\hat{x},\hat{w})$ 
and expand fields 
as follows, 
\begin{equation} 
\hat{a}(\hat{x},\hat{w})=\sum_{r_{1}=0}^{n}
\sum_{r_{2}} 
a_{r_{1}r_{2}}
\hat{Y}_{r_{1}r_{2}}(\hat{x},\hat{w}), 
\end{equation} 
If we consider a function corresponding to 
the above matrix, 
\begin{equation} 
a(x,w)=\sum_{r_{1}=0}^{n}\sum_{r_{2}}
a_{r_{1}r_{2}}
Y_{r_{1}r_{2}}(x,w), 
\end{equation} 
a product of the fields becomes noncommutative 
and associative. 
We would like to emphasize that 
$n$ plays a role of a cutoff parameter for the 
angular momentum $r_{1}$. 
$n$ is related to the size of the matrix 
and it gives UV and IR cut-off parameters. 
This means that 
a field theory on this noncommutative space 
has no divergences. 

When we consider the noncommutative field theory, 
an adjoint action of $\hat{G}_{\mu}$ becomes 
the following derivative operator
\begin{equation}
ad\left(\hat{G}_{\mu} \right)\rightarrow -2i\left(
w_{\mu\nu}\frac{\partial}{\partial x_{\nu}}
- x_{\nu}\frac{\partial}{\partial w_{\mu\nu}}
\right), 
\end{equation}
and an adjoint action of $\hat{G}_{\mu\nu}$ becomes
\begin{equation}
ad\left(\hat{G}_{\mu\nu} \right)\rightarrow 2\left(
x_{\mu}\frac{\partial}{\partial x_{\nu}}
-x_{\nu}\frac{\partial}{\partial x_{\mu}}
-w_{\mu\lambda}\frac{\partial}{\partial w_{\lambda\nu}}
+w_{\nu\lambda}\frac{\partial}{\partial w_{\lambda\mu}}
\right).
\label{adjointGmunu}
\end{equation}
We next discuss the Laplacian. 
It is natural to consider 
$ad(\hat{G}_{\mu})^{2}$ 
to be the Laplacian from the matrix model point of 
view \cite{ykfoursphere}. 
We now investigate the spectrum of the Laplacian. 
It is calculated as follows, 
\begin{equation}
\frac{1}{4}[\hat{G}_{\mu},[\hat{G}_{\mu},
\hat{Y}_{r_{1}r_{2}}]]
=\left(r_{1}(r_{1}+3)-r_{2}(r_{2}+1)\right)
\hat{Y}_{r_{1}r_{2}}. 
\label{eigenvalueofadG^{2}}
\end{equation}
It is well-known that 
noncommutative geometry is realized by the 
guiding center coordinates of electrons in 
a constant magnetic field. 
The authors in \cite{zhanghu} 
studied the quantum Hall system 
on a four dimensional sphere. 
Their system is composed of particles moving on a four dimensional 
sphere under an $SU(2)$ gauge field. 
The existence of Yang's $SU(2)$ monopole \cite{Yang} 
in the system makes the coordinates of the particles 
noncommutative. 
They showed that the configuration space of this system is 
locally $S^{4}\times S^{2}$ 
where $S^{2}$ represents an iso-spin space. 
We find that 
the eigenvalues (\ref{eigenvalueofadG^{2}}) 
exactly coincide with those of the Hamiltonian 
of a single particle on a four-sphere 
under the monopole background. 


\subsection{Contraction of fuzzy four-sphere algebra}
\label{sec:contractionfoursphere}
\hspace{0.4cm}
In this subsection, 
we consider the In\"{o}n\"{u}-Wigner 
contraction of 
the algebra (\ref{so(5,1)}) 
or (\ref{fourspherealgebras}) 
in the vicinity of the north pole. 
By virtue of the $SO(5)$ symmetry, this discussion is 
without loss of generality. 

$\hat{G}_{\mu}$ is decomposed to $\hat{G}_{5}\simeq n$ 
and $\hat{G}_{a}$, and 
$\hat{G}_{\mu\nu}$ to $\hat{G}_{ab}$ and $\hat{G}_{a5}$ 
at the north pole. 
Indices $a,b,\ldots$ run over 
$1$ to $4$.  
We rescale the matrices as 
\begin{equation}
\hat{G}_{a}^{\prime}=\frac{1}{\sqrt{n}}\hat{G}_{a}, 
\hspace{0.3cm}
\hat{G}_{ab}^{\prime}=\frac{1}{\sqrt{n}}\hat{G}_{ab}, 
\hspace{0.3cm}
\hat{G}_{a5}^{\prime}=\frac{1}{\sqrt{n}}\hat{G}_{a5}. 
\end{equation}
The radius of the four-sphere in the rescaled coordinate is 
\begin{equation}
\rho^{\prime 2}= 
\hat{x}_{i}^{\prime}\hat{x}_{i}^{\prime}
=\alpha^{2}\frac{n(n+4)}{n}=\frac{1}{n}\rho^{2}
\sim \alpha^{2}n .
\end{equation} 
To contract the algebra, we take 
$\rho^{\prime} \rightarrow \infty$ 
(or $n\rightarrow \infty$) limit with 
$\alpha$ fixed. 
From (\ref{gmunugnu=gmu}), we have 
\begin{equation} 
\hat{G}_{a5}=\frac{1}{n}\left(
4\hat{G}_{a}-\hat{G}_{ab}\hat{G}_{b}\right). 
\label{g5ga} 
\end{equation} 
If we use this equation, $\hat{G}_{a5}$ is written in terms of 
$\hat{G}_{a}$ and $\hat{G}_{ab}$ at the north pole. 
Therefore
independent matrices are now $\hat{G}_{a}$ and $\hat{G}_{ab}$. 
The commutation relations of them 
are given by  
\begin{eqnarray}
&&[\hat{G}_{a}^{\prime},\hat{G}_{b}^{\prime}]=
\frac{2}{\sqrt{n}}\hat{G}_{ab}^{\prime}, \cr
&&[\hat{G}_{a}^{\prime},\hat{G}_{bc}^{\prime}]=\frac{2}{\sqrt{n}}
\left(\delta_{ac}\hat{G}_{b}^{\prime}
-\delta_{ab}\hat{G}_{c}^{\prime}
 \right), \cr 
&&[\hat{G}_{ab}^{\prime},\hat{G}_{cd}^{\prime}]=
\frac{2}{\sqrt{n}}\left(
\delta_{bc}\hat{G}_{ad}^{\prime}+\delta_{ad}\hat{G}_{bc}^{\prime}
-\delta_{ac}\hat{G}_{bd}^{\prime}-\delta_{bd}\hat{G}_{ac}^{\prime}
\right)
. 
\end{eqnarray} 
 
Here we investigate the algebra of the fuzzy two-sphere 
in the large $n$ limit. 
It determines the noncommutativity for the 
coordinates of the four-sphere. 
One of the three coordinates of 
the fuzzy two-sphere is diagonalized 
as follows, 
\begin{equation}
\hat{N}_{3}=diag(n,n-2, \ldots ,-n+2,-n). 
\end{equation}
When we take the large $n$ limit, the two-brane charge 
no longer vanishes. 
It is because the contributions 
from the north pole and the south pole 
of the two-sphere do decouple. 
Then $\hat{N}_{3}$ takes the value $+n$ (or $-n$) 
in the large $n$ limit. 
Since the magnitudes of $\hat{N}_{1}$ and 
$\hat{N}_{2}$ are $O(1)$, 
the noncommutativity 
$\hat{G}_{ab}^{\prime}$ become as follows 
after taking the large $n$ limit, 
\begin{eqnarray}
&&\hat{G}_{12}^{\prime}=i\sqrt{n}{\bf 1}\hspace{0.2cm}
(\mbox{or}\hspace{0.2cm}
 -i\sqrt{n}{\bf 1}), 
\hspace{0.5cm} 
\hat{G}_{34}^{\prime}=-i\sqrt{n}{\bf 1}\hspace{0.2cm}
(\mbox{or}\hspace{0.2cm}
 i\sqrt{n}{\bf 1}), \cr
&&\hat{G}_{ab}^{\prime}=O(1) \hspace{0.4cm}
\left( (a,b)\neq(1,2),(3,4) \right). 
\label{noncommutativityflat}
\end{eqnarray}
Note that 
$\hat{M}_{i}^{\prime}
\simeq 0$. 
We have obtained the six dimensional noncommutative space 
whose coordinates have the following commutation relations, 
\begin{equation}
[\hat{G}_{1}^{\prime},\hat{G}_{2}^{\prime}]=2i{\bf 1}, \hspace{0.3cm}
[\hat{G}_{3}^{\prime},\hat{G}_{4}^{\prime}]=-2i{\bf 1},\hspace{0.3cm}
[\hat{N}_{1}^{\prime},\hat{N}_{2}^{\prime}]=2i{\bf 1}, 
\label{6dimplane1}
\end{equation}
or 
\begin{equation}
[\hat{G}_{1}^{\prime},\hat{G}_{2}^{\prime}]=-2i{\bf 1}, \hspace{0.3cm}
[\hat{G}_{3}^{\prime},\hat{G}_{4}^{\prime}]=2i{\bf 1},\hspace{0.3cm}
[\hat{N}_{1}^{\prime},\hat{N}_{2}^{\prime}]=-2i{\bf 1}.
\label{6dimplane2}
\end{equation}
In the language of $D$-brane, a flat $D3$-brane 
expands along the extra two dimensional plane 
and effectively form six dimensional space. 
We easily find that 
the fuzzy two-sphere plays the role of 
the symplectic form. 
Due to the presence of it, the symmetry of the plane is 
not $SO(4)$ but $SO(2)\times SO(2)$. 

We next study an action of noncommutative gauge theory 
on the noncommutative plane. 
Although we have just obtained 
the six dimensional noncommutative plane, 
the two dimensional space is an extra dimensional space 
and we obtain a four dimensional noncommutative gauge 
theory. 
The matrices $A_{\mu}$ are expanded as 
\begin{eqnarray}
A_{a}&=&\hat{x}_{a}+ \alpha \rho \hat{a}_{a} \cr 
&\equiv& \alpha \rho^{\prime}D^{\prime}_{a}, \cr
A_{5}&=&\alpha\rho^{\prime}\hat{\phi}^{\prime},
\end{eqnarray} 
where we have rescaled the field as 
$\sqrt{n}\hat{a}_{\mu}=\hat{a}_{\mu}^{\prime}$. 
$ad(D_{a})$ is the covariant derivative 
on the flat background 
and is given by 
\begin{equation}
ad(D_{a})=B_{ab}
\frac{\partial}{\partial x_{b}^{\prime}}
+[a_{a}^{\prime},\cdot\hspace{0.2cm}]_{\star}, 
\end{equation}
where $B_{12}=2i$, $B_{34}=-2i$ and 
other components of $B_{ab}$ are zero. 
It must be noted that the fields have 
dependence on not only $\hat{G}_{\mu}$ 
but also $\hat{N}_{i}$. 
An action of the noncommutative 
Yang-Mills is obtained from an action of 
the matrix model as 
\begin{eqnarray}
S=-\frac{(\alpha\rho^{\prime})^{4}}{4g^{2}}Tr
\left(
[D^{\prime}_{a},D^{\prime}_{b}]
[D^{\prime}_{a},D^{\prime}_{b}]
+2
[D^{\prime}_{a},\phi^{\prime}]
[D^{\prime}_{a},\phi^{\prime}]
\right) \cr
-\frac{\alpha\rho^{\prime}}{2g^{2}}Tr
\left(
\bar{\psi}\Gamma^{a}
[D^{\prime}_{a},\psi]
+\bar{\psi}\Gamma^{5}
[\phi^{\prime},\psi]
\right).
\end{eqnarray}
After replacing the trace with the integral as 
\begin{equation}
\frac{1}{N}Tr \rightarrow 
\frac{1}{\frac{8}{3}\pi\rho^{\prime 4}
\cdot 4\pi \rho^{\prime 2}
}
\int d^{4}x^{\prime}d^{2}y^{\prime}, 
\end{equation}
we have the following action 
of the noncommutative field theory, 
\begin{eqnarray}
S=-
\frac{3N\alpha^{4}}{128\pi^{2}\rho^{\prime 2}
g^{2}}
\int d^{4}x^{\prime}d^{2}y^{\prime}
\left(
[D^{\prime}_{a},D^{\prime}_{b}]
[D^{\prime}_{a},D^{\prime}_{b}]
+2
[D^{\prime}_{a},\phi^{\prime}]
[D^{\prime}_{a},\phi^{\prime}]
\right) \cr
-
\frac{N\alpha}{64\pi^{2}\rho^{\prime 5}g^{2}
}
\int d^{4}x^{\prime}d^{2}y^{\prime}
\left(
\bar{\psi}\Gamma^{a}
[D^{\prime}_{a},\psi]
+\bar{\psi}\Gamma^{5}
[\phi^{\prime},\psi]
\right).
\end{eqnarray}
Note that 
the fields propagate only on the $x$ directions 
since there are no derivatives for 
$y$ coordinates. 
After performing the integral $d^{2}y^{\prime}$, 
we obtain a four dimensional gauge theory with 
gauge coupling $g_{YM}^{2}=32\pi g^{2}/3N\alpha^{4}$.


\section{Fuzzy $2k$-sphere }
\label{sec:Fuzzy2k-sphere}
\subsection{Matrix description of fuzzy $2k$-sphere}
\hspace{0.4cm}
In this section, we study a 
higher dimensional fuzzy sphere 
\cite{ramgoo,horamgooram} and 
see that a lower dimensional fuzzy sphere is attached 
at each point on the higher dimensional fuzzy sphere 
as in the case of the fuzzy four-sphere.  

Let us start with the 
following $SO(2k+1,1)$ algebra
\footnote{
$k=1$ case is special since 
$\hat{G}_{\mu\nu}$ is written by 
$\hat{G}_{\mu}$ using the three-rank anti-symmetric 
tensor. 
In this case $\hat{G}_{\mu}$ generate the $SO(3)$ 
algebra. 
Some arguments of this section is applied 
for $k=1$ case. 
} 
\begin{equation}
[\hat{J}_{MN},\hat{J}_{OP}]=2\left(
\eta_{NO}\hat{J}_{MP}
+\eta_{MP}\hat{J}_{NO}
-\eta_{MO}\hat{J}_{NP}
-\eta_{NP}\hat{J}_{MO}
\right), 
\label{so(2k+1,1)algebra}
\end{equation}
where $M,N,\ldots$ run over $0$ to $2k+1$ and 
$\eta_{MN}=diag(-1,1,\ldots,1)$. 
$N_{k}$ dimensional representation 
of the algebra 
is constructed from the 
$n$-fold symmetric tensor product of 
the $(2k+1)$ dimensional 
Gamma matrices as 
\begin{eqnarray}
&&\hat{J}_{0\mu}=\left(
\Gamma_{\mu}\otimes 1\otimes \cdots \otimes 1
+1\otimes \Gamma_{\mu} \otimes \cdots \otimes 1
+\cdots 
+1\otimes 1\otimes \cdots \otimes \Gamma_{\mu}
\right)_{Sym}, \cr
&&\hat{J}_{\mu\nu}=\left(
\Gamma_{\mu\nu}\otimes 1\otimes \cdots \otimes 1
+1\otimes \Gamma_{\mu\nu} \otimes \cdots \otimes 1
+\cdots 
+1\otimes 1\otimes \cdots \otimes \Gamma_{\mu\nu}
\right)_{Sym}. 
\end{eqnarray}
The value of 
$N_{k}$ is summarized in appendix \ref{sec:someformulae} 
and becomes $n^{\frac{k(k+1)}{2}}$ for a large $n$. 
We rewrite $\hat{J}_{0\mu}$ and 
$\hat{J}_{\mu\nu}$ 
as $\hat{G}_{\mu}$ and 
$\hat{G}_{\mu\nu}$, and 
they represent coordinates 
of a fuzzy $2k$-sphere and a fuzzy $2(k-1)$-sphere 
respectively as shown from now on. 
The commutation relations of them are given by 
\begin{eqnarray} 
[\hat{G}_{\mu},\hat{G}_{\nu}]&=&
2\hat{G}_{\mu\nu}, 
\cr
[\hat{G}_{\mu},\hat{G}_{\nu\lambda}]&=&2\left(
\delta_{\mu\nu}\hat{G}_{\lambda}-\delta_{\mu\lambda}
\hat{G}_{\nu}
\right), \cr
[\hat{G}_{\mu\nu},\hat{G}_{\lambda\rho}]&=&2\left(
\delta_{\nu\lambda}\hat{G}_{\mu\rho}
+\delta_{\mu\rho}\hat{G}_{\nu\lambda}
-\delta_{\mu\lambda}\hat{G}_{\nu\rho}
-\delta_{\nu\rho}\hat{G}_{\mu\lambda}
\right).
\label{so(2k+1,1)algebra2}
\end{eqnarray}
We can confirm that $\hat{G}_{\mu}$ 
satisfy 
the following relation, 
\begin{equation}
\epsilon^{\mu_{1}\cdots \mu_{2k}\mu_{2k+1}}
\hat{G}_{\mu_{1}} \cdots 
\hat{G}_{\mu_{2k}}
=C_{k}\hat{G}_{\mu_{2k+1}},  
\label{fuzzyrelation}
\end{equation}
where $C_{k}$ is a constant which depends on $n$ 
and $\epsilon^{\mu_{1}\cdots \mu_{2k+1}}$ 
is the $SO(2k+1)$ invariant tensor. 
This relation respects the $SO(2k+1)$ symmetry 
which is the isometry of a $2k$-sphere. 
The radius of the fuzzy $2k$-sphere is given by 
\begin{equation}
\hat{x}_{\mu}\hat{x}_{\mu}
=\alpha^{2}\hat{G}_{\mu}\hat{G}_{\mu}
=\alpha^{2}n(n+2k)\equiv \rho^{2}.
\label{fuzzy2ksphereradius}
\end{equation}

We now show that 
a fuzzy $2(k-1)$-sphere exists at each point on 
a fuzzy $2k$-sphere 
and coordinates of the fuzzy $2(k-1)$-sphere 
are given 
by $\hat{G}_{ab}$ $(a,b=1,\ldots,2k)$. 
We consider the algebra 
at the north pole of the fuzzy $2k$-sphere, 
$\hat{G}_{2k+1} \simeq n $.
$\hat{G}_{ab}$ are generators of the $SO(2k)$ 
rotation around the north pole. 
The lower dimensional fuzzy sphere is embedded in 
the $SO(2k)$ algebra. 
If we regard the vector coordinates of 
the $2(k-1)$-sphere 
as $i\hat{G}_{\alpha\hspace{0.05cm}2k}$, 
$i\hat{G}_{\alpha\hspace{0.05cm}2k}$ 
and $\hat{G}_{\alpha\beta}$ 
$(\alpha,\beta=1,\ldots,2k-1)$ 
form 
$SO(2k-1,1)$ algebra. 
This algebra indeed describes 
a fuzzy $2(k-1)$-sphere. 
$\hat{G}_{a\hspace{0.05cm}2k+1}$ depend 
on $\hat{G}_{a}$ and $\hat{G}_{ab}$
since 
they are written as 
\begin{equation}
\hat{G}_{a\hspace{0.05cm}2k+1}=
\frac{1}{n}\left(2k\hat{G}_{a}
-\hat{G}_{ab}\hat{G}_{b}\right), 
\end{equation}
where we used the formula (\ref{gmunugnu=gmu}).
Since the fuzzy $2k$-sphere has the $SO(2k+1)$ symmetry, 
the fuzzy $2(k-1)$-sphere 
exists at each point 
on the fuzzy $2k$-sphere. 

It can be proved that the fuzzy $2(k-1)$-sphere 
on the fuzzy $2k$-sphere satisfies 
the relations which are analogous to 
(\ref{fuzzyrelation}) and 
(\ref{fuzzy2ksphereradius}). 
To prove them, we use the following identifications 
\begin{equation}
\hat{G}_{\alpha\beta}^{(2k)}=
\hat{G}_{\alpha\beta}^{(2(k-1))}, 
\hspace{0.4cm}
i\hat{G}_{\alpha\hspace{0.05cm}2k}^{(2k)}
=\hat{G}_{\alpha}^{(2(k-1))}. 
\label{identify2kand2k-2}
\end{equation}
We begin with the relation (\ref{fuzzyrelation}) 
for the fuzzy $2k$-sphere, and set 
$\hat{G}_{2k+1}=n $, 
\begin{eqnarray}
C_{k}n&=&
\epsilon^{\mu_{1}\cdots \mu_{2k}\hspace{0.05cm}2k+1}
\hat{G}_{\mu_{1}}^{(2k)} \cdots 
\hat{G}_{\mu_{2k}}^{(2k)}  \cr
&=&\epsilon^{\mu_{1}\cdots \mu_{2k} \hspace{0.05cm}2k+1}
\hat{G}_{\mu_{1}\mu_{2}} ^{(2k)}
\cdots  
\hat{G}_{\mu_{2k-1}\mu_{2k} }^{(2k)} \cr
&=&2k\epsilon^{\mu_{1}\cdots \mu_{2k-1}
\hspace{0.05cm}2k \hspace{0.05cm}2k+1}
\hat{G}_{\mu_{1}\mu_{2}}^{(2k)}\cdots  
\hat{G}_{\mu_{2k-1}\hspace{0.05cm}2k }^{(2k)} \cr
&=&-i2k\epsilon^{\mu_{1}\cdots \mu_{2k-1}
\hspace{0.05cm}2k \hspace{0.05cm}2k+1}
\hat{G}_{\mu_{1}\mu_{2}}^{(2(k-1))}\cdots  
\hat{G}_{\mu_{2k-1} }^{(2(k-1))} \cr
&=&-i2k\epsilon^{\mu_{1}\cdots \mu_{2k-1}
\hspace{0.05cm}2k \hspace{0.05cm}2k+1}
\hat{G}_{\mu_{1}}^{(2(k-1))}\cdots 
\hat{G}_{\mu_{2k-1} }^{(2(k-1))}. 
\end{eqnarray}
The identifications 
(\ref{identify2kand2k-2}) are used 
from the third line to the forth line. 
We have arrived at the following equation, 
\begin{equation}
\epsilon^{\mu_{1}\cdots \mu_{2k-1}}
\hat{G}_{\mu_{1}}^{(2(k-1))}\cdots  
\hat{G}_{\mu_{2k-1} }^{(2(k-1))}=\frac{i}{2k}nC_{k}. 
\label{2k-2relationfrom2k1}
\end{equation}
This expression is equivalent to the following 
expression 
(see appendix \ref{sec:someformulae}),  
\begin{equation}
\epsilon^{\mu_{1}\cdots \mu_{2k-2}\mu_{2k-1}}
\hat{G}_{\mu_{1}} \cdots 
\hat{G}_{\mu_{2k-2}}
=C_{k-1}\hat{G}_{\mu_{2k-1}}. 
\label{fuzzyrelation2k-2}
\end{equation}
Then we have obtained the relation 
(\ref{fuzzyrelation2k-2}) for the fuzzy $2(k-1)$-sphere 
from the relation 
(\ref{fuzzyrelation}) for the fuzzy $2k$-sphere 
using the identifications (\ref{identify2kand2k-2}). 
Let us now calculate the radius of the fuzzy $2(k-1)$-sphere  
using the relations for the fuzzy $2k$-sphere. 
From the formulae given in 
appendix \ref{sec:someformulae}, 
we can calculate as 
\begin{eqnarray}
2kn(n+2k)&=&\hat{G}_{\mu\nu}^{(2k)}
\hat{G}_{\nu\mu}^{(2k)} \cr 
&=&\hat{G}_{ab}^{(2k)}\hat{G}_{ba}^{(2k)}
+2\hat{G}_{2k+1\hspace{0.05cm}a}^{(2k)}
\hat{G}_{a\hspace{0.05cm}2k+1}^{(2k)} \cr 
&=&\hat{G}_{ab}^{(2k)}\hat{G}_{ba}^{(2k)}
+2\left(n(n+2k)-\hat{G}_{2k+1}^{(2k)}
\hat{G}_{2k+1}^{(2k)} \right). 
\end{eqnarray}
Then we have 
\begin{eqnarray}
\hat{G}_{ab}^{(2k)}\hat{G}_{ba}^{(2k)}
&=&(2k-2)n(n+2k)+2\hat{G}_{2k+1}^{(2k)}
\hat{G}_{2k+1}^{(2k)} \cr 
&=&2kn(n+2k-2), 
\label{lowdimensionradius1}
\end{eqnarray}
where we used $\hat{G}_{2k+1}^{(2k)}=n$. 
$\hat{G}_{ab}\hat{G}_{ba}$ is also calculated as 
\begin{eqnarray}
\hat{G}_{ab}^{(2k)}\hat{G}_{ba}^{(2k)}
&=& \hat{G}_{\alpha\beta}^{(2k)}\hat{G}_{\beta\alpha}^{(2k)} 
+2\hat{G}_{2k\hspace{0.05cm}\alpha}^{(2k)}
\hat{G}_{\alpha\hspace{0.05cm}2k}^{(2k)}
\cr 
&=& \hat{G}_{\alpha\beta}^{(2(k-1))}
\hat{G}_{\beta\alpha}^{(2(k-1))} 
+2\hat{G}_{\alpha}^{(2(k-1))}
\hat{G}_{\alpha}^{(2(k-1))}
 \cr
&=&2k\hat{G}_{\alpha}^{(2(k-1))} 
\hat{G}_{\alpha}^{(2(k-1))}.
\label{lowdimensionradius2}
\end{eqnarray}
From the second line to the third line, 
we have used the following equation 
which is derived from (\ref{GmuGmu=nn}) 
and 
(\ref{GmunuGmunu=nn}), 
\begin{equation}
\hat{G}_{\alpha\beta}^{(2(k-1))} 
\hat{G}_{\beta\alpha}^{(2(k-1))}  
=(2k-2)\hat{G}_{\alpha}^{(2(k-1))} 
\hat{G}_{\alpha}^{(2(k-1))} .
\end{equation}
Therefore
we find from (\ref{lowdimensionradius1}) and 
(\ref{lowdimensionradius2}) 
that the fuzzy $2(k-1)$-sphere has a radius 
$\alpha^{2}n(n+2k-2)$. 
The size of the extra dimensions is also given by $\rho$. 

Eigenvalues of the matrices represent 
the positions of quanta in the noncommutative space-time, 
and we have $N_{k}\sim n^{\frac{k(k+1)}{2}}$ quanta 
in this system. 
For the $k=3$ case, we have $N_{3}\sim n^{6}$ quanta. 
Since a fuzzy four-sphere exists at each points 
on a six-sphere, 
there are $n^{3}$ points on the six-sphere 
and $N_{2}\sim n^{3}$ points on the fuzzy four-sphere. 
Since the fuzzy four-sphere has 
a fuzzy two-sphere at each point, 
$n^{3}$ can be divided into $n^{2}\cdot n$. 
For the $k=4$ case, 
$N_{4}\sim n^{10}=n^{4}\cdot n^{3} \cdot n^{2}\cdot n$ 
where each factor represents 
the number of the quanta on an eight-sphere, a six-sphere, 
a four-sphere and a two-sphere. 
The area of the unit quantum on a $2k$-sphere 
is given by 
\begin{equation}
\frac{2\pi^{k+\frac{1}{2}}\rho^{2k}}
{\Gamma(k+\frac{1}{2})}\frac{N_{k-1}}{N_{k}}
\sim \frac{\rho^{2k}}{n^{k}} 
\sim \alpha^{2k}n^{k}, 
\label{plancklength2ksphere}
\end{equation}
where $\frac{2\pi^{k+\frac{1}{2}}}
{\Gamma(k+\frac{1}{2})}\rho^{2k}$ is the area of a $2k$-sphere. 
This is a noncommutative scale on this system.

We began with (\ref{so(2k+1,1)algebra}) or 
(\ref{so(2k+1,1)algebra2}) for the 
fuzzy $2k$-sphere algebra. 
On the other hand, 
we may use the $SO(2k+1)$ algebra as the definition of 
the fuzzy $2k$-sphere algebra thanks to 
the isomorphic relation 
between $SO(2k+2)/U(k+1)$ and 
$SO(2k+1)/U(k)$. 
This relation also helps us to guess 
the relationship between 
the fuzzy $2k$-sphere 
and the fuzzy $2(k-1)$-sphere 
as follows. 
If we start from the 
coset space $SO(2k+1)/U(k)$, 
it is written as 
\begin{equation}
SO(2k+1)/U(k)
= SO(2k+1)/SO(2k) \times 
SO(2k-1)/U(k-1). 
\end{equation}
This shows that a fuzzy $2(k-1)$-sphere 
exists at each point on a $2k$-sphere. 

We can show that lower dimensional brane charges 
in the fuzzy $2k$-branes vanish. 
Let us first consider the $k=3$ case. 
A two-brane charge 
$tr[\hat{x}_{\mu},\hat{x}_{\nu}]$ vanishes 
because of the cyclicity of the trace for 
finite matrices. 
As for a four-brane charge, 
we consider at the north pole for 
simplicity, and 
use the identification (\ref{identify2kand2k-2}) 
for $k=3$: 
\begin{equation}
\hat{G}_{\alpha\beta}^{(6)}=\hat{G}_{\alpha\beta}^{(4)}, 
\hspace{0.4cm}
i\hat{G}_{\alpha6}^{(6)}=\hat{G}_{\alpha}^{(4)}. 
\label{so(6)inso(7,1)}
\end{equation}
A four-brane charge is given by a 
completely anti-symmetrized product of four 
matrices as 
$Tr\hat{x}_{[a}\hat{x}_{b}\hat{x}_{c}\hat{x}_{d]}
=\alpha^{4}
Tr\hat{G}_{[a}\hat{G}_{b}\hat{G}_{c}\hat{G}_{d]}$.
When the indices do not include $6$, 
it becomes as 
\begin{eqnarray}
&&\hspace{0.4cm}\epsilon^{abcde}
Tr[\hat{G}_{a}^{(6)},\hat{G}_{b}^{(6)}]
[\hat{G}_{c}^{(6)},\hat{G}_{d}^{(6)}]
\cr
&&=\epsilon^{abcde}
Tr[\hat{G}_{a}^{(4)},\hat{G}_{b}^{(4)}]
[\hat{G}_{c}^{(4)},\hat{G}_{d}^{(4)}]
\cr
&&=\epsilon^{abcde}
Tr\hat{G}_{a}^{(4)}\hat{G}_{b}^{(4)}
\hat{G}_{d}^{(4)}\hat{G}_{d}^{(4)}
\cr
&&=Tr\hat{G}_{e}^{(4)}=0. 
\end{eqnarray}
We have used the identification (\ref{so(6)inso(7,1)}) 
from the first line to the second line. 
When the indices include $6$, 
we can calculate as 
\begin{eqnarray}
&&\hspace{0.4cm}\epsilon^{abcde}
Tr[\hat{G}_{a}^{(6)},\hat{G}_{b}^{(6)}]
[\hat{G}_{c}^{(6)},\hat{G}_{6}^{(6)}]
\cr
&&=-2i\epsilon^{abcde}
Tr[\hat{G}_{a}^{(4)},\hat{G}_{b}^{(4)}]
\hat{G}_{c}^{(4)}
\cr
&&=-2i\epsilon^{abcde}
Tr\hat{G}_{a}^{(4)}\hat{G}_{b}^{(4)}
\hat{G}_{c}^{(4)} \cr
&&\propto Tr[\hat{G}_{d}^{(4)},\hat{G}_{e}^{(4)}]
=0.
\end{eqnarray}
Therefore the four-brane charge vanishes. 
We have shown that lower dimensional brane charges 
vanish at each point on the fuzzy six-sphere. 
They do not vanish after the contraction of the 
algebra (see the next subsection). 
Results for the $k=4$ case 
are analogous. 

Noncommutative field theories on the fuzzy spheres 
can be defined through matrix models by the same way as
in section \ref{section:NCGT}. 
Fields are expanded by the noncommutative spherical harmonics. 
It is characterized 
by the irreducible representations of $SO(2k+1)$ 
which are represented by the Young diagram of the 
low length $(r_{1},r_{2},\ldots,r_{k})$ 
where $n\ge r_{1}\ge r_{2}\ge\cdots\ge r_{k}$, 
and 
summing up the dimensions of all irreducible representations 
leads 
to the square of $N_{k}$ \cite{ramgoo}. 
{\it Coordinates} which correspond to 
the representation with $r_{k}(k\ge 3)\neq 0$ are written 
in terms of $\hat{G}_{a}$ and $\hat{G}_{ab}$ 
at each point on the fuzzy $2k$-sphere. 
Then the fuzzy $2k$-sphere is a  
$k(k+1)$ dimensional space and 
it has an extra $k(k-1)$ dimensional space 
which forms the fuzzy $2(k-1)$-sphere. 

Before finishing this subsection, 
let us comment on a large $N$ limit 
which gives a smooth manifold. 
A classical manifold is recovered by 
a large $N$ limit with the radius $\rho$ fixed. 
In the limit, the coordinates 
$\hat{x}_{\mu}$ and $\hat{w}_{\mu\nu}$ 
become commuting matrices. 
Since the coordinates $\hat{w}_{\mu\nu}$ 
originate from the lower dimensional sphere, 
we obtain the classical manifold 
$S^{2k}\times S^{2(k-1)} \times \cdots \times S^{2}$. 


\subsection{Contraction of fuzzy $2k$-sphere algebra}
\hspace{0.4cm}
The contraction of the fuzzy $2k$-sphere 
algebra is considered in this subsection. 
It is shown that there appear a $k(k+1)$ dimensional 
noncommutative plane 
including $k(k-1)$ extra dimensions 
which originate from the lower dimensional 
fuzzy-sphere. 
We first consider the $k=3$ case. 
We rescale the generators as 
$\hat{G}_{a}^{\prime}=\frac{1}{\sqrt{n}}\hat{G}_{a}$ 
and  
$\hat{G}_{ab}^{\prime}=\frac{1}{\sqrt{n}}\hat{G}_{ab}$ 
$(a,b=1,\ldots,6)$ for later convenience. 
The commutation relations between $\hat{G}_{a}^{\prime}$ 
become 
\begin{equation}
[\hat{G}_{a}^{\prime},\hat{G}_{b}^{\prime}]
=\frac{2}{\sqrt{n}}\hat{G}_{ab}^{\prime}. 
\end{equation}
The $SO(6)$ generators $\hat{G}_{ab}$ are 
identified with the generators 
of the fuzzy four-sphere algebra as in 
(\ref{so(6)inso(7,1)}). 
$\hat{G}_{5}^{(4)}$ is diagonalized as 
\begin{equation}
\hat{G}_{5}^{(4)}=diag(n,n-2,\cdots,-n+2,-n). 
\end{equation}
We consider the region 
in the vicinity of the north pole of the four-sphere, 
$\hat{G}_{5}^{(4)}\simeq n$ 
and 
take a large $n$ limit to obtain 
a infinitely extended plane.
We can also diagonalize 
$\hat{N}_{3}^{(4)}
=-\frac{i}{4}\eta^{3}_{ab}\hat{G}_{ab}^{(4)}$ 
which is defined in (\ref{thooftsymbol}). 
In the region, the generators become as 
$\hat{G}_{a}^{(4)\prime}\simeq O(1)$, 
$\hat{N}_{i}^{(4)\prime}\simeq O(1)$ 
and 
$\hat{N}_{3}^{(4)\prime}\simeq \sqrt{n}$. 
Referring to the discussions in section 
\ref{sec:contractionfoursphere}, 
we obtain the following 
twelve dimensional noncommutative plane: 
\begin{eqnarray}
&&[\hat{G}_{1}^{\prime},\hat{G}_{2}^{\prime}]
=\frac{2}{\sqrt{n}}\hat{G}_{12}^{(4)\prime}
=
2i{\bf 1},
\hspace{0.4cm}
[\hat{G}_{3}^{\prime},\hat{G}_{4}^{\prime}]
=\frac{2}{\sqrt{n}}\hat{G}_{34}^{(4)\prime }
= -2i{\bf 1}, \cr
&&[\hat{G}_{5}^{\prime},\hat{G}_{6}^{\prime}]
=\frac{-2i}{\sqrt{n}}\hat{G}_{5}^{(4)\prime}
= -2i{\bf 1},
\cr 
&&[\hat{G}_{16}^{\prime},\hat{G}_{26}^{\prime}]
=-[\hat{G}_{1}^{(4)\prime},\hat{G}_{2}^{(4)\prime}]
=-2i{\bf 1},
\hspace{0.4cm}
[\hat{G}_{36}^{\prime},\hat{G}_{46}^{\prime}]
=-[\hat{G}_{3}^{(4)\prime},\hat{G}_{4}^{(4)\prime}]
=2i{\bf 1}, \cr
{\rule[-2mm]{0mm}{9mm}\ }    
&&[\hat{G}_{13}^{\prime},\hat{G}_{23}^{\prime}]
=[\hat{G}_{13}^{(4)\prime},\hat{G}_{23}^{(4)\prime}]
=-[\hat{N}_{1}^{(4)\prime},\hat{N}_{2}^{(4)\prime}]
=-2i{\bf 1}.
\label{twelvenoncommutativeplane}
\end{eqnarray}
Other commutation relations are zero. 
The forth and fifth commutation relations 
come from the fuzzy four-sphere 
and the last one from the fuzzy two-sphere. 
If we apply the same procedure to the $k=4$ case, 
we obtain a twenty dimensional noncommutative plane 
with twelve extra dimensions. 
We find that the lower dimensional sphere 
gives the symplectic structure and 
the symmetry of the noncommutative plane is 
$SO(2)^{k}$. 



\section{Fuzzy sphere as classical solution of 
matrix model}
\label{sec:fuzzysphereandmatrixmodel}
\hspace{0.4cm}
We consider matrix model realizations of 
the fuzzy spheres in this section. 
Original matrix models \cite{BFSS,IKKT} do not 
have static higher dimensional fuzzy spheres 
as classical solutions. 
Let us consider IIB matrix model \cite{IKKT} 
as a example, 
\begin{equation}
S=-\frac{1}{g^{2}}Tr
\left(\frac{1}{4}
[A_{\mu},A_{\nu}][A_{\mu},A_{\nu}]
+\frac{1}{2}\bar{\psi}\Gamma^{\mu}
[A_{\mu},\psi]
\right).
\label{IIBaction}
\end{equation}
Supersymmetries of this model are given by 
\begin{eqnarray}
&&\delta^{(1)} \psi = \frac{i}{2} 
  \left[ A_{\mu} ,A_{\nu}\right] \Gamma^{\mu\nu} \epsilon,  
   \hspace{0.4cm}
\delta^{(1)} A_{\mu} = i\bar{\epsilon}
     \Gamma^{\mu}\psi,  \cr 
&&\delta^{(2)} \psi = \xi,  \hspace{2.7cm}
\delta^{(2)} A_{\mu} = 0, 
\end{eqnarray}
and these form ${\cal N}=2$ supersymmetry algebra. 

In the matrix model, eigenvalues of 
bosonic variables are interpreted as space-time 
coordinates \cite{AIKKT}. 
Space-time coordinates are represented  
by matrices and noncommutative geometry 
is expected to appear. 
A flat noncommutative space 
which is given by the following 
is a formal classical solution 
of this model in a large $N$ limit, 
\begin{equation}
[\hat{x}_{\mu},\hat{x}_{\nu}]=iC_{\mu\nu}{\bf 1} 
\end{equation}
and this solution preserves the ${\cal N}=1$ 
supersymmetry. 
We can embed a noncommutative manifold whose 
isometry is a subgroup of $SO(10)$ in the matrix model. 
The fuzzy spheres satisfy this condition. 
Since the fuzzy $2k$-spheres satisfy the following 
equation 
\begin{eqnarray}
[\hat{G}_{\nu},[\hat{G}_{\mu},\hat{G}_{\nu}]]
=2[\hat{G}_{\nu},\hat{G}_{\mu\nu}]
=-16\hat{G}_{\mu} 
=-\frac{16}{C_{k}}
\epsilon^{\mu_{1}\cdots\mu_{2k}\mu}
\hat{G}_{\mu_{1}}\cdots\hat{G}_{\mu_{2k}},
\end{eqnarray}
the fuzzy $2k$-sphere becomes a classical 
solution of the matrix model 
by adding a ($2k+1$)-rank Chern-Simons term 
or a mass term. 
A Chern-Simons term 
\footnote{
Note that $C_{k} (k\neq 2)$ depend on $n$ while 
$C_{2}$ does not. 
The equations of motion 
determine the dimension of the fuzzy sphere 
for $k\neq 2$.
} 
appeared in a low energy effective action of $D$-branes 
under a R-R background \cite{myers}. 
A matrix model with a mass term is 
investigated in \cite{yk2}. 
See also \cite{azumabagnoud} in which 
a massive matrix model is analyzed form the viewpoint of 
a supermatrix model. 
At present it is not clear whether 
the matrix model receives such deformations. 
The issue should be discussed in the future publications. 
The embedding of the curved space-time in matrix models 
is also discussed in 
\cite{IKTW,yk2,ykfoursphere,kabattaylor,
nairdaemi,IK,kitazawahomogeneous}. 

Although the fuzzy sphere solutions 
do not preserve the supersymmetry, 
it is shown to be recovered locally on the sphere. 
(It can be easily checked that 
(\ref{6dimplane1}) and 
(\ref{twelvenoncommutativeplane}) preserve 
the ${\cal N}=1$ supersymmetry.) 
As for the fuzzy two-sphere, 
we can construct matrix models 
in which the fuzzy two-sphere is 
a supersymmetric classical 
solution \cite{IKTW,BMN}. 
In these deformed models, the supersymmetry transformations 
are deformed such that the fuzzy two-sphere is supersymmetric 
and form a ${\cal N}=2$ algebra in spite of the deformation. 

The advantage of the matrix model construction of 
noncommutative gauge theories is that 
the gauge symmetry is manifest. 
If we expand $A_{\mu}$ around a classical background as 
\begin{equation}
A_{\mu}=\hat{x}_{\mu}+\hat{a}_{\mu},
\end{equation}
the fluctuation $\hat{a}_{\mu}$ behaves as a gauge field 
on a noncommutative background $\hat{x}_{\mu}$. 
The gauge symmetry in noncommutative 
gauge theories originates from 
the unitary symmetry of the matrix model, and  
$A_{\mu}$ is a gauge covariant quantity. 
By considering such a expansion, 
the relationship between matrix models and 
field theories becomes more manifest. 
One can introduce a natural ultraviolet cut-off 
which is given by the size of matrix into 
noncommutative Yang-Mills theories, 
and letting it large in a way 
leads to usual field theories. 
Fluctuations of different backgrounds 
in the matrix model 
are described by different gauge theories.  
From this point of view, 
$A_{\mu}$ are clearly background independent 
\cite{seiberg0008013}. 



\section{Dielectric effect and 
higher dimensional fuzzy sphere}
\label{sec:Dielectriceffectandfuzzysphere}
\hspace{0.4cm}
In this section, 
we show that $n^{\frac{k(k-1)}{2}}$ 
spherical $D(2k-1)$-branes give 
a fuzzy $2k$-sphere in the matrix model. 
The existence of lower dimensional fuzzy spheres 
is understood from the dielectric effect. 
The first point is established by calculating the value of the 
action (\ref{IIBaction}) for fuzzy spheres. 
This calculation is analogous to the calculation 
for flat $D$-branes in \cite{chepelev}. 
We identify $g^{2}$ as $g_{s}l_{s}^{\hspace{0.1cm}4}$ 
in this calculation, where 
$g_{s}$ and $l_{s}$ are the string coupling constant 
and the string length scale respectively \cite{IKKT}. 
It seems reasonable to suppose that 
the string length scale is related to 
the noncommutative scale on 
the fuzzy sphere. 
(See, for example, \cite{IIKKstringscale} on this point.) 
From (\ref{plancklength2ksphere}) we can 
find the following relation, 
\begin{equation}
l_{s}^{2k}\sim \frac{\rho^{2k}}{n^{k}}
\sim \alpha^{2k}n^{k}. 
\end{equation}
The value of the action 
for a fuzzy sphere is estimated 
as follow, 
\begin{eqnarray}
S&=&-\frac{1}{4g_{s}l_{s}^{\hspace{0.1cm}4}}
Tr[A_{\mu},A_{\nu}][A_{\mu},A_{\nu}] \cr
&=&-\frac{\alpha^{4}}{g_{s}l_{s}^{\hspace{0.1cm}4}}
Tr\hat{G}_{\mu\nu}\hat{G}_{\mu\nu} \cr 
&=&\frac{\alpha^{4}}{g_{s}l_{s}^{\hspace{0.1cm}4}}
N_{k}2kn(n+2k) \cr 
&\sim&\frac{1}{g_{s}l_{s}^{\hspace{0.1cm}2k}}
\rho^{2k}N_{k-1}.  
\end{eqnarray}
where $g_{s}l_{s}^{\hspace{0.1cm}2k}$ is the tension 
of a $D(2k-1)$-brane up to a numerical coefficient. 
$\sim$ means that we considered a large $n$. 
This is 
the energy of $N_{k-1}$ spherical 
$D(2k-1)$-branes, where $N_{k-1}\sim n^{\frac{k(k-1)}{2}}$. 
We can conclude that a fuzzy $2k$-sphere is comprised 
of $N_{k-1}$ spherical $D(2k-1)$-branes. 
On the other hand, we showed that 
a fuzzy $2k$-sphere has a fuzzy $2(k-1)$-sphere 
at each point. 
The consistency of these two viewpoints 
becomes clear by 
considering the dielectric effect. 
We may say that $N_{k-1}$ $D$-instantons are present 
at each point on the $D(2k-1)$-branes 
since we are considering $N_{k-1}$ 
$D(2k-1)$-branes. 
Since 
$D$-branes expand into a fuzzy sphere under 
a constant anti-symmetric tensor background due to 
the dielectric effect \cite{myers}, 
the $N_{k-1}$ $D$-instantons expand into 
a fuzzy $2(k-1)$-sphere. 
It must be noted that 
coordinates of the fuzzy $2(k-1)$-sphere 
are embedded in {\it extra dimensions}. 
It is known that 
matrix descriptions of $D$-branes 
correspond to $D$-branes 
under NS-NS or R-R form backgrounds. 
$D$-instantons under a constant 
R-R three-form field strength background 
expand to a fuzzy two-sphere. 
In general, $D$-instantons under a R-R $(2k+1)$-form 
field strength background 
expand to a fuzzy $2k$-sphere. 
A fuzzy $2(k-1)$-sphere appears at each point 
on the fuzzy $2k$-sphere since 
a constant $(2k+1)$-form contains 
lower dimensional forms. 
This section can be summarized in the following sentences: 
{\it A fuzzy $2k$-sphere is given by a 
bound state of $N_{k-1}$ $D(2k-1)$-branes 
and $N_{k}$ $D$-instantons. 
The dielectric effect occurs at each point 
on the $D(2k-1)$-branes, and 
this effect makes extra spherical spaces. } 


\section{Summary and Discussions}
\label{sec:summaryanddiscussions}
\hspace{0.4cm}
In this paper, we studied matrix descriptions 
of higher dimensional sphererical branes. 
A higher dimensional sphere which is described 
by the coset space $SO(2k+1)/SO(2k)$ 
can not be quantized 
since it does not admit a symplectic structure. 
Instead of considering this space, 
we consider the coset space $SO(2k+1)/U(k)$. 
It is a K\"{a}hler space and 
a symplectic structure 
is given by a K\"{a}hler form \cite{masuda}. 
We showed that 
there is a fuzzy $2(k-1)$-sphere 
at each point on the fuzzy $2k$-sphere. 
The fuzzy $2k$-sphere is a $k(k+1)$ dimensional space 
including $k(k-1)$ {\it extra dimensions}. 
The coordinates 
of the fuzzy $2k$-sphere and the fuzzy $2(k-1)$-sphere 
which exists at each point on the fuzzy $2k$-sphere 
are given by 
$\hat{G}_{\mu}$ and $\hat{G}_{\mu\nu}$ 
respectively. 
The lower dimensional spheres serve 
the symplectic structure. 
This fact becomes more obvious after 
the contraction of the algebra as discussed 
in section \ref{sec:contractionfoursphere}. 
It is well known that 
the quantum Hall system is an example of noncommutative 
geometry. 
As is discussed in \cite{zhanghu}, 
the quantum Hall system on a four dimensional sphere 
is constructed by considering 
a system of particles under a $SU(2)$ gauge field. 
The existence of the gauge field background 
produces the noncommutativity, and 
the same observation applies to 
the fuzzy sphere system. 
The existence of the lower dimensional fuzzy sphere 
produces the noncommutativity on the higher dimensional 
sphere. 

The values of the matrix model for fuzzy spheres 
are computed 
in section \ref{sec:Dielectriceffectandfuzzysphere}. 
The computation shows that 
the fuzzy $2k$-sphere system is composed of $N_{k-1}$ 
$D(2k-1)$-branes where $N_{k-1}$ corresponds to 
the dimension of the fuzzy $2(k-1)$-sphere. 
The fuzzy $2k$-sphere is formed from $N_{k}$ 
$D$-instantons 
under a constant R-R $(2k+1)$-form field strength background. 
This is the so called Myers effect. 
Since the $(2k+1)$-form includes a $(2k-1)$-form, 
the fuzzy $2(k-1)$-sphere appears on the fuzzy $2k$-sphere. 
This is also due to the Myers effect. 
The appearance of the fuzzy $2(k-1)$-sphere is 
closely related to the overlap of $N_{k-1}$ 
$D(2k-1)$-branes. 
The Myers effect arises in such a interesting way 
for higher dimensional spheres. 

It is known that a matrix model around a noncommutative 
background gives the definition of a noncommutative 
Yang-Mills theory on the background. 
A noncommutative Yang-Mills on a fuzzy sphere 
is obtained through a matrix model. 
Since we take account of fluctuations 
only around $\hat{G}_{\mu}$, 
we would obtain a $2k$ dimensional noncommutative 
Yang-Mills. 
The extra $k(k-1)$ dimensional space plays the same role 
as the gauge field background in the quantum Hall effect.

The advantage of compact noncommutative manifolds is that 
one can construct them in terms of finite size 
matrices 
while a solution which represents a noncommutative plane 
cannot be represented by finite size matrices. 
From the viewpoint of noncommutative field theories, 
$N$ plays the role of the cutoff parameter. 
The noncommutative field theories 
are completely finite as long as 
$N$ is finite. 
We showed that 
a gauge theory on a noncommutative plane 
were reproduced from 
a gauge theory on 
a fuzzy sphere by taking a large $N$ limit.  
The existence of the symplectic form breaks 
the full Lorentz invariance of the plane and 
we obtaine a noncommutative plane which 
is the Heisenberg algebra. 

\newpage

\vspace{1.0cm}
\begin{center}
{\bf Acknowledgments}
\end{center}
\hspace{0.4cm}
I gratefully acknowledge helpful discussions 
with Y. Shibusa on several points in this paper. 
I also thank S. Iso, Y. Kitazawa, T. Masuda 
and T. Tada 
for useful discussions. 


\renewcommand{\theequation}{\Alph{section}.\arabic{equation}}
\appendix
\section{Some formulae}
\label{sec:someformulae}
\setcounter{equation}{0} 
\hspace{0.4cm}
In this appendix, we summarize several formulae 
involving 
$\hat{G}_{\mu}$ and $\hat{G}_{\mu\nu}$ 
in diverse dimensions. 
$\hat{G}_{\mu}$ and $\hat{G}_{\mu\nu}$ 
are generators of $SO(2k+1,1)$, 
and they are constructed from 
the $n$-fold symmetrized tensor product 
of the Gamma matrices. 
The dimensions $N_{k}$ of the representation 
are given by 
\begin{eqnarray}
&&N_{1}=n+1, \hspace{0.4cm}
N_{2}=\frac{1}{6}(n+1)(n+2)(n+3), \cr 
&&N_{3}=\frac{1}{360}(n+1)(n+2)(n+3)^{2}(n+4)(n+5), \cr
&&N_{4}=\frac{1}{302400}(n+1)(n+2)(n+3)^{2}
(n+4)^{2}(n+5)^{2}(n+6)(n+7). 
\end{eqnarray}
These are calculated in \cite{ramgoo}. 
We have the following invariants 
\begin{equation}
\hat{G}_{\mu}\hat{G}_{\mu}=n(n+2k)  
\label{GmuGmu=nn}
\end{equation}
and 
\begin{equation}
\hat{G}_{\mu\nu}\hat{G}_{\nu\mu}=2kn(n+2k). 
\label{GmunuGmunu=nn}
\end{equation} 
The following relations are also satisfied 
\begin{equation}
\hat{G}_{\mu\nu}\hat{G}_{\nu}
=2k\hat{G}_{\mu}
\label{gmunugnu=gmu}
\end{equation}
and 
\begin{equation}
\hat{G}_{\mu\nu}\hat{G}_{\nu\lambda} 
=n(n+2k)\delta_{\mu\lambda}
 +(k-1)\hat{G}_{\mu}\hat{G}_{\lambda}
 -k\hat{G}_{\lambda}\hat{G}_{\mu}.
\end{equation}
$G_{\mu}$ satisfy the following relation 
\begin{equation}
\epsilon^{\mu_{1}\cdots\mu_{2k}\mu_{2k+1}}
\hat{G}_{\mu_{1}}
\cdots\hat{G}_{\mu_{2k}}
=C_{k}\hat{G}_{\mu_{2k+1}}
\label{appnoncommurelationG}
\end{equation}
where $\epsilon^{\mu_{1}\mu_{2}\cdots\mu_{2k}\mu_{2k+1}}$ 
is the $SO(2k+1)$ invariant tensor. 
$C_{k}$ is a constant which depends on $n$,   
\begin{equation}
C_{1}=2i, \hspace{0.2cm}
C_{2}=8(n+2), \hspace{0.2cm}
C_{3}=-48i(n+2)(n+4), \hspace{0.2cm}C_{4}=-384(n+2)(n+4)(n+6).
\label{valueofC}
\end{equation}
The details of this calculation are given in 
\cite{azumabagnoud}. 
By multiplying the equation (\ref{appnoncommurelationG}) 
by $\hat{G}_{\mu_{2k+1}}$, 
we have 
\begin{eqnarray}
\epsilon^{\mu_{1}\cdots \mu_{2k}\mu_{2k+1}}
\hat{G}_{\mu_{1}} \cdots 
\hat{G}_{\mu_{2k}}\hat{G}_{\mu_{2k+1}}
&=&C_{k}n(n+2k) \cr  
&=&\frac{i}{2(k+1)}nC_{k+1}, 
\label{noncommurelationG2}
\end{eqnarray}
where we have used the following relation 
which is found from (\ref{valueofC}), 
\begin{equation}
C_{k}
=-i2k(n+2k-2)C_{k-1}. 
\label{valuesofCk}  
\end{equation}

\section{Notations of Gamma matrices}
\label{sec:gammmamatrices}
\setcounter{equation}{0} 
\hspace{0.4cm}
We summarize an 
explicit representation of 
Gamma matrices in odd dimensions $d=2k+1$. 
Higher dimensional ones are composed of lower 
dimensional ones as 
\begin{eqnarray} 
\Gamma_{\mu}  
 &=&   \left( \begin{array}{c c}
  0 & -i\gamma_{\mu} \\ 
 i\gamma_{\mu}  & 0   \\
 \end{array} \right)=\gamma_{\mu}\otimes \sigma_{2}
  \hspace{0.2cm}(\mu=1,\ldots,2k-1),     \cr
\Gamma_{d-1}  
 &=&   \left( \begin{array}{c c}
  0 & 1_{2^{k-1}} \\ 
  1_{2^{k-1}}  & 0   \\
 \end{array} \right) ={\bf 1_{2^{k-1}}}\otimes \sigma_{1},     \cr
\Gamma_{d}  
 &=&   \left( \begin{array}{c c}
  1_{2^{k-1}} & 0 \\ 
 0  & -1_{2^{k-1}}   \\
 \end{array} \right)={\bf 1_{2^{k-1}}}\otimes \sigma_{3},        
\end{eqnarray}
where $\sigma_{\mu}$ is the Pauli matrices. 
They satisfy the Clifford algebra, 
\begin{equation}
\{\Gamma_{\mu} ,\Gamma_{\nu} \}=2\delta_{\mu\nu}
\hspace{0.2cm}(\mu,\nu=1,\ldots,d=2k+1).
\end{equation}



\end{document}